\pdfoutput=1
\documentclass[
  reprint,
  superscriptaddress,
 amsmath,amssymb,
 aps,
 prl
]{revtex4-1}

\usepackage{graphicx}
\usepackage{dcolumn}
\usepackage{bm}
\usepackage{color}

\begin{document}

\title{
  Symmetry-protected difference between
  spin Hall and anomalous Hall effects of
  a periodically driven multiorbital metal
}

\author{Naoya Arakawa}
\email{arakawa@phys.chuo-u.ac.jp}
\affiliation{The Institute of Science and Engineering,
  Chuo University, Bunkyo, Tokyo, 112-8551, Japan}
\author{Kenji Yonemitsu}
\affiliation{The Institute of Science and Engineering,
  Chuo University, Bunkyo, Tokyo, 112-8551, Japan}
\affiliation{Department of Physics,
  Chuo University, Bunkyo, Tokyo 112-8551, Japan}


\begin{abstract}
  Nonequilibrium quantum states can be controlled
  via the driving field in periodically driven systems. 
  Such control, which is called Floquet engineering,
  has opened various phenomena, such as the light-induced anomalous Hall effect.  
  There are expected to be some essential differences
  between the anomalous Hall and spin Hall effects of periodically driven systems
  because of the difference in time-reversal symmetry.  
  However, these differences remain unclear
  due to the lack of Floquet engineering of the spin Hall effect.
  Here we show that 
  when the helicity of circularly polarized light is changed
  in a periodically driven $t_{2g}$-orbital metal,  
  the spin current generated by the spin Hall effect remains unchanged, 
  whereas the charge current generated by the anomalous Hall effect is reversed.
  This difference is protected by the symmetry of a time reversal operation.
  Our results offer a way
  to distinguish the spin current and charge current via light
  and could be experimentally observed in pump-probe measurements of
  periodically driven Sr$_{2}$RuO$_{4}$.
\end{abstract}
\maketitle


\hspace{-13pt}
\textbf{Introduction}

Periodically driven systems enable
the realization of various nonequilibrium quantum states
and their control.
Periodically driven systems are realized by a time-periodic field, 
and their properties in a nonequilibrium steady state 
can be described by
the Floquet theory~\cite{Floquet1,Floquet2,Floquet-review1,Floquet-review2,Floquet-review3},
in which the effective Hamiltonian is independent of time.
In fact,
various theoretical predictions,
such as the light-induced anomalous Hall effect (AHE)~\cite{Oka-PRB,Light-AHE-exp1,Light-AHE-exp2}
and the Floquet time crystal~\cite{TimeCry-theory,TimeCry-exp1,TimeCry-exp2},
are confirmed by experiments.
Then,
since the effective Hamiltonian of the Floquet theory
depends on parameters of the driving field,
its properties can be controlled by tuning the driving field.
This is called Floquet engineering~\cite{Floquet-review1,Floquet-review2,Floquet-review3}.
For example,
it is possible to change 
the magnitude, sign, and bond anisotropy of exchange interactions
of Mott insulators~\cite{Exc-Mott1,Exc-Mott2,Exc-Mott3,Exc-Mott4,Exc-Mott5}.
The Floquet engineering has been studied in many fields of physics,
including condensed-matter, cold-atom, and optical physics. 

Although there are many studies of the AHE of periodically driven systems,
the Floquet engineering of the spin Hall effect (SHE) is still lacking.
The SHE is
the key phenomenon in spintronics~\cite{InvSHE1,InvSHE2,SSE1,SSE2,Bauer-review}.
In the SHE,
an electron spin current, a flow of the spin angular momentum,
is generated by an electric field perpendicular to it~\cite{SHE-theory,SHE-exp,SHE-review}.
This is the spin version of the AHE, in which
an electron charge current is generated~\cite{Karplus-Luttinger,AHE-review}.
A significant difference between the AHE and SHE
is about time-reversal symmetry (TRS):
TRS is broken in the AHE, whereas it holds in the SHE.
Since TRS can be broken by circularly polarized light~\cite{Chirality},
there should be some essential differences between 
those of a periodically driven system.
It is highly desirable to investigate
the intrinsic SHE of a periodically driven multiorbital metal
because the intrinsic SHE, the SHE intrinsic to the electronic structure,
can be engineered by the driving field
and several multiorbital metals, such as Pt,
have the huge SHE~\cite{Kon-SHE-Pt,Kon-SHE-Ru}.

Here we show that
in a multiorbital metal driven by circularly polarized light, 
the charge current generated by the AHE can be reversed 
by changing the helicity of light, 
whereas the spin current generated by the SHE remains unchanged.
This is demonstrated by constructing a theory of pump-probe measurements
of the AHE and SHE
of a periodically driven $t_{2g}$-orbital metal coupled to a heat bath
and evaluating their conductivities numerially.
This significant difference between the AHE and SHE
results from the difference in TRS
and thus should hold in many periodically driven systems.
We also show that 
spin-orbit coupling (SOC) is vital for the SHE of the periodically driven multiorbital metal,
whereas it is unnecessary for the AHE. 
This property is distinct from that of non-driven metals. 

\hspace{-13pt}
\textbf{Results and Discussion}

\hspace{-13pt}
\textbf{Periodically driven $t_{2g}$-orbital metal}

\begin{figure*}
  \includegraphics[width=170mm]{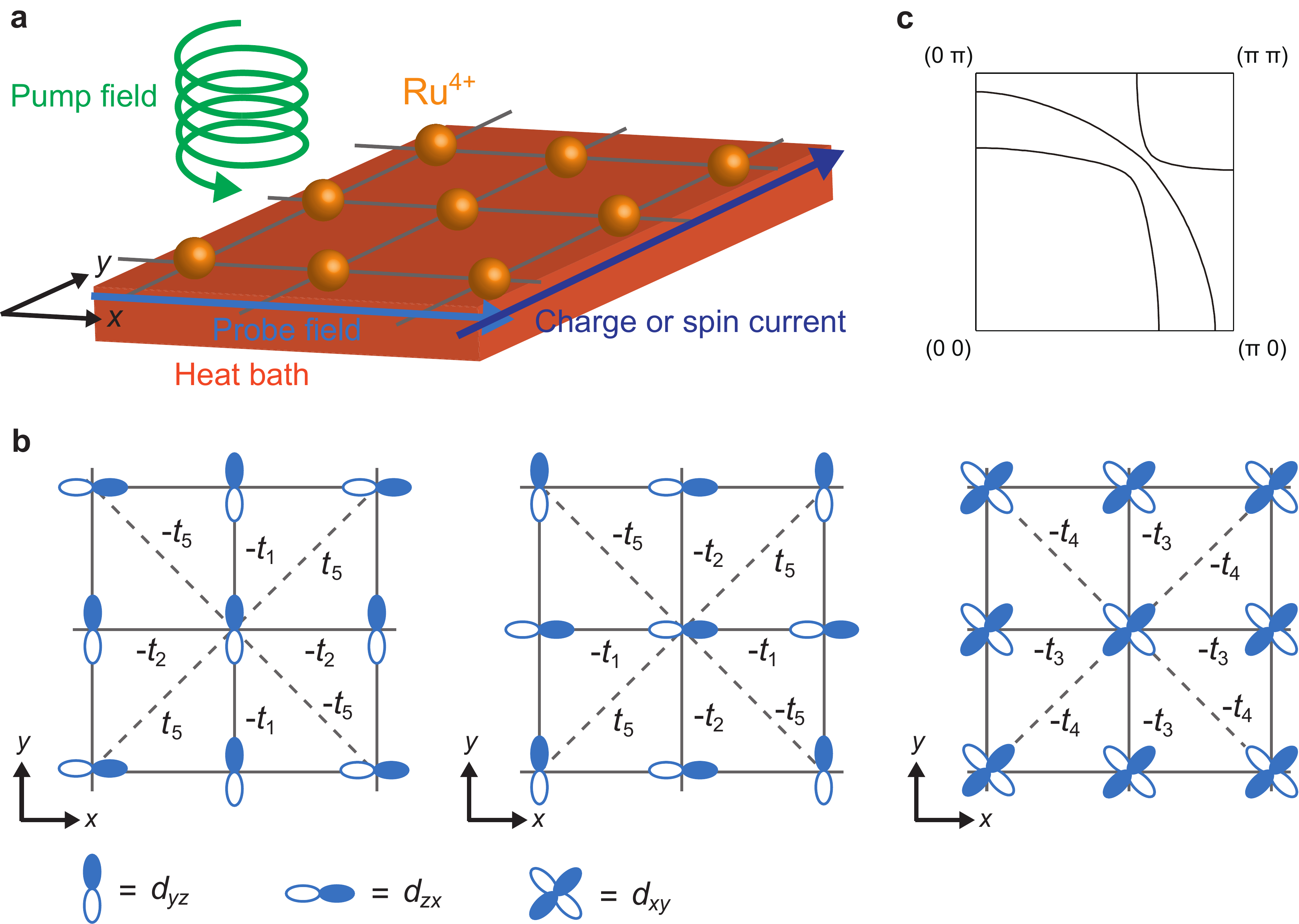}
  \caption{\label{fig1}
    \textbf{Set-up of the anomalous Hall or spin Hall effect and electronic properties of our model.}
    \textbf{a}, Set-up of the anomalous Hall or spin Hall effect
    for our model of Sr$_{2}$RuO$_{4}$
    driven by circularly polarized light in the presence of the coupling to a heat bath.
    In Sr$_{2}$RuO$_{4}$, Ru ions form the square lattice;
    at each ion, four electrons occupy
    the Ru $t_{2g}$ orbitals (i.e., the $d_{yz}$, $d_{zx}$, and $d_{xy}$ orbitals).
    In the pump-probe measurements of the anomalous Hall and spin Hall effects,
    the probe field induces 
    the charge and spin currents, respectively, perpendicular to it,
    and
    the pump field,
    a field of left- or right-circularly polarized light,
    periodically drives Sr$_{2}$RuO$_{4}$.
    The nonequilibrium steady state is realized because of the coupling to the heat bath. 
    \textbf{b}, 
    The finite hopping processes of electrons in $t_{2g}$ orbitals
    on the square lattice.
    The $d_{yz}$, $d_{zx}$, and $d_{xy}$ represent these orbitals.
    $t_{1}$, $t_{2}$, and $t_{3}$
    are the nearest-neighbor hopping integrals, and 
    $t_{4}$ and $t_{5}$ are the next nearest-neighbor ones.
    \textbf{c},
    The Fermi surface obtained for the non-driven case of our model
    in the quarter of the Brillouin zone.
    The other parts are reproducible by using the rotational symmetry. 
  }
\end{figure*}

We consider a $t_{2g}$-orbital metal coupled to a heat bath
in the presence of a field $\mathbf{A}(t)$ (Fig. \ref{fig1}a):  
\begin{align}
  H(t)=H_{\textrm{s}}(t)+H_{\textrm{sb}}+H_{\textrm{b}}.\label{eq:Htot}
\end{align}
(Note that in the $t_{2g}$-orbital metal, such as Sr$_{2}$RuO$_{4}$, 
  electrons occupy the $t_{2g}$ orbitals, i.e., the $d_{yz}$, $d_{zx}$, and $d_{xy}$ orbitals.)
First, 
$H_{\textrm{s}}(t)$ is the system Hamiltonian, 
the Hamiltonian of $t_{2g}$-orbital electrons with $\mathbf{A}(t)$,
\begin{align}
  \hspace{-10pt}
  H_{\textrm{s}}(t)
  &=\sum_{\mathbf{k}}\sum_{a,b=d_{yz},d_{zx},d_{xy}}\sum_{\sigma,\sigma^{\prime}=\uparrow,\downarrow}
  \bar{\epsilon}_{ab}^{\sigma\sigma^{\prime}}(\mathbf{k},t)
  c_{\mathbf{k} a\sigma}^{\dagger}c_{\mathbf{k} b\sigma^{\prime}}.\label{eq:Hs}
\end{align}
Here
$c_{\mathbf{k} a\sigma}^{\dagger}$ and $c_{\mathbf{k} a\sigma}$
are the creation and annihilation operators, respectively,
of an electron for orbital $a$ with momentum $\mathbf{k}$ and spin $\sigma$,
and 
\begin{align}
  \bar{\epsilon}_{ab}^{\sigma\sigma^{\prime}}(\mathbf{k},t)
  =[\epsilon_{ab}(\mathbf{k},t)-\mu\delta_{a,b}]\delta_{\sigma,\sigma^{\prime}}
  +\xi_{ab}^{\sigma\sigma^{\prime}},\label{eq:ebar}
\end{align}
where  
$\epsilon_{ab}(\mathbf{k},t)$,
$\mu$, and $\xi_{ab}^{\sigma\sigma^{\prime}}$ are
the kinetic energy with the Peierls phase factors due to $\mathbf{A}(t)$,
the chemical potential,
and SOC, respectively (see Methods).
Throughout this paper, we use the unit $\hbar=1$, $k_{\textrm{B}}=1$,
and $a_{\textrm{lc}}=1$,
where $a_{\textrm{lc}}$ is the lattice constant.
In addition to $H_{\textrm{s}}(t)$,
we have considered $H_{\textrm{b}}$ and $H_{\textrm{sb}}$,
the Hamiltonian of
a B\"{u}ttiker-type heat bath~\cite{HeatBath1,HeatBath2,Tsuji,Mikami}
at temperature $T_{\textrm{b}}$
and the system-bath coupling Hamiltonian (see Methods).
This is 
because 
a nonequilibrium steady state can be realized
due to the damping coming from
the second-order perturbation of $H_{\textrm{sb}}$~\cite{Tsuji,Keldysh}. 

The parameters of $H_{\textrm{s}}(t)$ are chosen 
to reproduce the electronic structure of Sr$_{2}$RuO$_{4}$~\cite{Sr2RuO4-review}.
The hopping integrals on the square lattice are parametrized by 
$t_{1}$, $t_{2}$, $t_{3}$, $t_{4}$, and $t_{5}$
(Fig. \ref{fig1}b)~\cite{NA-Ru-model}, 
and $\mu$ is determined from the condition $n_{\textrm{e}}=4$,
where $n_{\textrm{e}}$ is the electron number per site;
the value of $\mu$ is fixed at that determined in the non-driven case. 
We set
$(t_{1},t_{2},t_{3},t_{4},t_{5})=(0.675,0.09,0.45,0.18,0.03)$ (eV)~\cite{NA-Ru-model}
and $\xi=0.17$ eV~\cite{SOC-Oguchi}, where $\xi$ is the coupling constant of SOC,
in order that the Fermi surface (Fig. \ref{fig1}c) is consistent with
that observed experimentally~\cite{Sr2RuO4-ARPES}. 

\hspace{-13pt}
\textbf{Theory of pump-probe measurements of the SHE and AHE}

The SHE and AHE of a periodically driven system
are detectable by pump-probe measurements.
In the pump-probe measurements~\cite{Opt-review},
a system is periodically driven by the pump field $\mathbf{A}_{\textrm{pump}}(t)$,
and its properties are analyzed by the probe field $\mathbf{A}_{\textrm{prob}}(t)$. 
Thus, 
we set $\mathbf{A}(t)=\mathbf{A}_{\textrm{pump}}(t)+\mathbf{A}_{\textrm{prob}}(t)$
and treat the effects of $\mathbf{A}_{\textrm{pump}}(t)$ in the Floquet theory
and those of $\mathbf{A}_{\textrm{prob}}(t)$
in the linear-response theory~\cite{Ecks-Trans,Mikami};
in our analyses, 
$\mathbf{A}_{\textrm{pump}}(t)$ is chosen to be
\begin{align}
  \mathbf{A}_{\textrm{pump}}(t)={}^{t}(A_{0}\cos\Omega t\ A_{0}\sin(\Omega t+\delta)),\label{eq:A_pump}
\end{align}
where $\Omega=2\pi/T$ and 
$T$ is the period of $\mathbf{A}_{\textrm{pump}}(t)$.
The anomalous-Hall and spin-Hall conductivities
$\sigma_{yx}^{\textrm{C}}(t,t^{\prime})$ and $\sigma_{yx}^{\textrm{S}}(t,t^{\prime})$
are defined as
\begin{align}
  \sigma_{yx}^{\textrm{Q}}(t,t^{\prime})
  &=\frac{1}{i\omega}
  \frac{\delta \langle j_{\textrm{Q}}^{y}(t)\rangle}{\delta A_{\textrm{prob}}^{x}(t^{\prime})},
  \label{eq:AHC-SHC-start}
\end{align}
where
$\langle j_{\textrm{C}}^{y}(t)\rangle$ and $\langle j_{\textrm{S}}^{y}(t)\rangle$
are the expectation values of the charge and spin current density operators,
respectively.
In our AHE or SHE,
we have considered the charge or spin current, respectively, generated along the $y$ axis
with the probe field applied along the $x$ axis (Fig. \ref{fig1}a).
(Note that our SHE is different from
  the SHE of light, in which the helicity-dependent transverse shift of light
  at an interface is induced~\cite{LSHE1,LSHE2,LSHE3}.)
Then,
the charge and spin current operators
$J_{\textrm{C}}^{y}(t)=Nj_{\textrm{C}}^{y}(t)$ and
$J_{\textrm{S}}^{y}(t)=Nj_{\textrm{S}}^{y}(t)$, where $N$ is the number of sites,
are determined from the continuity equations
(see Methods)~\cite{Mahan,Mizo,Kon-SHE-Pt,Kon-SHE-Ru}:
\begin{align}
  J_{\textrm{Q}}^{y}(t)
  =\sum_{\mathbf{k}}\sum_{a,b}\sum_{\sigma}v_{ab\sigma}^{(\textrm{Q})y}(\mathbf{k},t)
  c_{\mathbf{k} a\sigma}^{\dagger}(t)c_{\mathbf{k} b\sigma}(t),\label{eq:JC-JS}
\end{align}
where $v_{ab\sigma}^{(\textrm{C})y}(\mathbf{k},t)=(-e)\frac{\partial \epsilon_{ab}(\mathbf{k},t)}{\partial k_{y}}$,
$v_{ab\sigma}^{(\textrm{S})y}(\mathbf{k},t)=\frac{1}{2}\textrm{sgn}(\sigma)\frac{\partial \epsilon_{ab}(\mathbf{k},t)}{\partial k_{y}}$,
and
$\textrm{sgn}(\sigma)=1$ or $-1$ for $\sigma=\uparrow$ or $\downarrow$, respectively.
By combining Eq. (\ref{eq:JC-JS}) with Eq. (\ref{eq:AHC-SHC-start})
and using a method of Green's functions~\cite{Mahan,Keldysh,Keldysh-review,Kadanoff-Baym},
we can express $\sigma_{yx}^{\textrm{Q}}(t,t^{\prime})$ in terms of
electron Green's functions (see Methods).

To analyze the SHE and AHE in the nonequilibrium steady state,
we consider the time-averaged dc anomalous-Hall and spin-Hall conductivities
$\sigma_{yx}^{\textrm{C}}$ and $\sigma_{yx}^{\textrm{S}}$, 
\begin{align}
  \sigma_{yx}^{\textrm{Q}}
  =\lim_{\omega\rightarrow 0}\textrm{Re}\int_{0}^{T}\frac{dt_{\textrm{av}}}{T}
  \int_{-\infty}^{\infty}dt_{\textrm{rel}}e^{i\omega t_{\textrm{rel}}}
  \sigma_{yx}^{\textrm{Q}}(t,t^{\prime}),\label{eq:t-av-sig}
\end{align}
where 
$t_{\textrm{rel}}=t-t^{\prime}$ and $t_{\textrm{av}}=(t+t^{\prime})/2$~\cite{Mikami}.
Since we can calculate Eq. (\ref{eq:t-av-sig})
in a way similar to that for charge transport of
single-orbital systems~\cite{Ecks-Trans,Mikami,Tsuji},
we present the final result here (for the derivation, see Supplementary Note 1):
\begin{align}
  &\sigma_{yx}^{\textrm{Q}}
  =\frac{1}{N}\sum_{\mathbf{k}}\sum_{a,b,c,d}\sum_{\sigma,\sigma^{\prime}}
  \int_{-\Omega/2}^{\Omega/2}\frac{d\omega^{\prime}}{2\pi}
  \sum_{m,l,n,q=-\infty}^{\infty}
  [v_{ab\sigma}^{(\textrm{Q})y}(\mathbf{k})]_{ml}\notag\\
  &\times [v_{cd\sigma^{\prime}}^{(\textrm{C})x}(\mathbf{k})]_{nq}
  \Bigl\{  
  \frac{\partial [G_{b\sigma c\sigma^{\prime}}^{\textrm{R}}(\mathbf{k},\omega^{\prime})]_{ln}}
       {\partial \omega^{\prime}}
    [G_{d\sigma^{\prime}a\sigma}^{<}(\mathbf{k},\omega^{\prime})]_{qm}\notag\\
    &\ \ \ \ \ \ \ \ \ \ \
    -[G_{b\sigma c\sigma^{\prime}}^{<}(\mathbf{k},\omega^{\prime})]_{ln}
    \frac{\partial [G_{d\sigma^{\prime}a\sigma}^{\textrm{A}}(\mathbf{k},\omega^{\prime})]_{qm}}
         {\partial \omega^{\prime}}
    \Bigr\},\label{eq:sig^Q}
\end{align}
where $[v_{ab\sigma}^{(\textrm{Q})\nu}(\mathbf{k})]_{mn}$ (Q $=$ C or S, $\nu=y$ or $x$)
and $[G_{a\sigma b\sigma^{\prime}}^{r}(\mathbf{k},\omega^{\prime})]_{mn}$ ($r=$ R, A, or $<$)
are given by
\begin{align}
  [v_{ab\sigma}^{(\textrm{Q})\nu}(\mathbf{k})]_{mn}
  &=\int_{0}^{T}\frac{dt}{T}e^{i(m-n)\Omega t}v_{ab\sigma}^{(\textrm{Q})\nu}(\mathbf{k},t),\\
  [G_{a\sigma b\sigma^{\prime}}^{r}(\mathbf{k},\omega^{\prime})]_{mn}
  &=\int_{-\infty}^{\infty}dt_{\textrm{rel}}e^{i(\omega^{\prime}+\frac{m+n}{2}\Omega)t_{\textrm{rel}}}
  \int_{0}^{T}\frac{dt_{\textrm{av}}}{T}\notag\\
  &\times e^{i(m-n)\Omega t_{\textrm{av}}}
  G_{a\sigma b\sigma^{\prime}}^{r}(\mathbf{k};t,t^{\prime}),
\end{align}
respectively;
the three Green's functions are determined from the Dyson equation 
with the damping $\Gamma$ due to the system-bath coupling (see Methods).
(For the energy dispersion of our model,
  see Supplementary Note 2.)

\hspace{-13pt}
\textbf{Helicity-independent $\sigma_{yx}^{\textrm{S}}$ and helicity-dependent $\sigma_{yx}^{\textrm{C}}$}

\begin{figure*}
  \includegraphics[width=180mm]{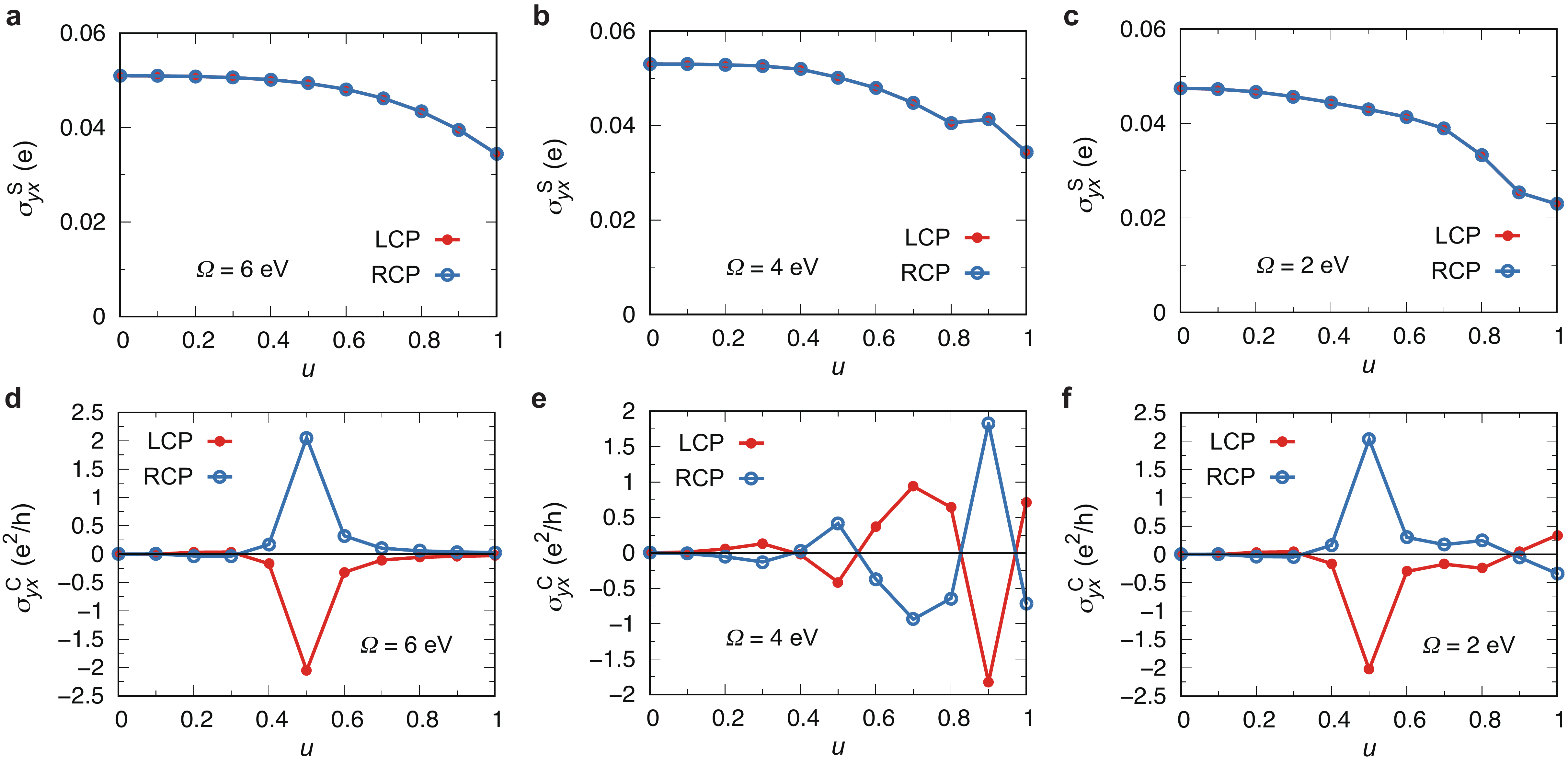}
  \caption{\label{fig2}
    \textbf{Helicity dependences of the spin Hall and anomalous Hall conductivities.}
    \textbf{a}, \textbf{b}, \textbf{c},
    The dependences of the spin Hall conductivity $\sigma_{yx}^{\textrm{S}}$
    on the dimensionless quantity $u=eA_{0}$
    in the case of left- or right-circularly polarized light (LCP or RCP)
    at $\Omega=6$, $4$, and $2$ eV,
    where $\Omega$ is the frequency of light. 
    The red and blue curves correspond to those
    in the case of left- or right-circularly polarized light, respectively. 
    \textbf{d}, \textbf{e}, \textbf{f},
    The dependences of the anomalous Hall conductivity $\sigma_{yx}^{\textrm{C}}$ on $u=eA_{0}$
    in the case of left- or right-circularly polarized light
    at $\Omega=6$, $4$, and $2$ eV.
    The same notations as those in \textbf{a}, \textbf{b}, \textbf{c} are used. 
  }
\end{figure*}

We evaluate $\sigma_{yx}^{\textrm{C}}$ and $\sigma_{yx}^{\textrm{S}}$ numerically.
(For the details of the numerical calculations, see Methods.)
We set $\Gamma=0.03$ eV and $T_{\textrm{b}}=0.05$ eV;
$\Gamma$ is chosen to be smaller than $T_{\textrm{b}}$
because
the system is supposed to be well described by the Fermi liquid. 
To study how $\sigma_{yx}^{\textrm{C}}$ and $\sigma_{yx}^{\textrm{S}}$
are affected by the helicity of light, 
we consider
the $\mathbf{A}_{\textrm{pump}}(t)$'s for $\delta=0$ and $\pi$ [Eq. (\ref{eq:A_pump})],
$\mathbf{A}_{\textrm{LCP}}(t)$ and $\mathbf{A}_{\textrm{RCP}}(t)$, 
which correspond to
the cases of 
the left- and right-circularly polarized light, respectively.
We show how 
  $\sigma_{yx}^{\textrm{S}}$ and $\sigma_{yx}^{\textrm{C}}$ depend on
  a dimensionless quantity $u=eA_{0}=eE_{0}/\Omega$.
  Note that the $u$ dependence at fixed $\Omega$
  corresponds to the dependence on $E_{0}$, the amplitude of the electric field.

$\sigma_{yx}^{\textrm{S}}$ and $\sigma_{yx}^{\textrm{C}}$ have
the different helicity dependences. 
Figure \ref{fig2}a shows the dependence of $\sigma_{yx}^{\textrm{S}}$ on $u=eA_{0}$
for $\mathbf{A}_{\textrm{pump}}(t)=\mathbf{A}_{\textrm{LCP}}(t)$ or $\mathbf{A}_{\textrm{RCP}}(t)$
at $\Omega=6$ eV.
The $\sigma_{yx}^{\textrm{S}}$ for $\mathbf{A}_{\textrm{pump}}(t)=\mathbf{A}_{\textrm{LCP}}(t)$
is the same as that for $\mathbf{A}_{\textrm{pump}}(t)=\mathbf{A}_{\textrm{RCP}}(t)$.
This property holds even at $\Omega=4$ and $2$ eV (Figs. \ref{fig2}b and \ref{fig2}c).
Note that $\Omega=6$, $4$, and $2$ eV correspond to
$\Omega > W$, $\Omega\approx W$, and $\Omega < W$, respectively,
where $W(\approx 4\ \textrm{eV})$ is the bandwidth in the non-driven case.
Meanwhile,
$\sigma_{yx}^{\textrm{C}}$'s
for $\mathbf{A}_{\textrm{pump}}(t)=\mathbf{A}_{\textrm{LCP}}(t)$ and $\mathbf{A}_{\textrm{RCP}}(t)$
are opposite in sign and the same in magnitude
at $\Omega=6$, $4$, and $2$ eV (Figs. \ref{fig2}d{--}\ref{fig2}f).
Although such helicity-dependent $\sigma_{yx}^{\textrm{C}}$
  was experimentally shown in graphene~\cite{Light-AHE-exp2},
  its origin may be unexplored.
Note that
  the difference between the $u$ dependences of $\sigma_{yx}^{\textrm{S}}$ and $\sigma_{yx}^{\textrm{C}}$
  can be qualitatively understood
  by considering the dominant terms of the Bessel functions due to the Peierls phase factors
  (see Supplementary Note 3 and Supplementary Figure 1).

\begin{figure*}
  \includegraphics[width=174mm]{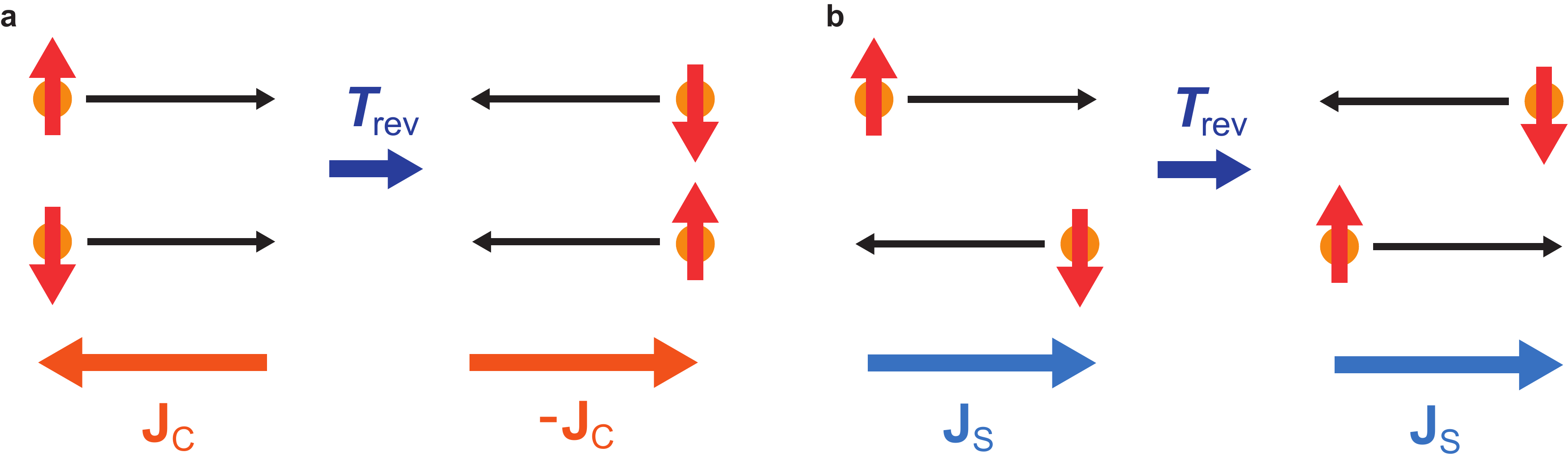}
  \caption{\label{fig3}
    \textbf{Time-reversal symmetry of the charge current and spin current.}
    \textbf{a}, \textbf{b},
    The charge currents and the spin currents 
    before and after the time-reversal operation $T_{\textrm{rev}}$. 
    The charge current $\mathbf{J}_{\textrm{C}}$ and
    the spin current $\mathbf{J}_{\textrm{S}}$ are 
    $\mathbf{J}_{\textrm{C}}=(-e)(\mathbf{J}_{\uparrow}+\mathbf{J}_{\downarrow})$
    and $\mathbf{J}_{\textrm{S}}=(1/2)(\mathbf{J}_{\uparrow}-\mathbf{J}_{\downarrow})$,
    where $\mathbf{J}_{\uparrow}$ and $\mathbf{J}_{\downarrow}$ are 
    the spin-up and spin-down electron currents, respectively.
    As a result of $T_{\textrm{rev}}$,
    $\mathbf{J}_{\uparrow}$ and $\mathbf{J}_{\downarrow}$
    become $-\mathbf{J}_{\downarrow}$ and $-\mathbf{J}_{\uparrow}$, respectively. 
    Thus, 
    $\mathbf{J}_{\textrm{C}}$ changes its sign (\textbf{a}),
    whereas $\mathbf{J}_{\textrm{S}}$ remains the same (\textbf{b}).
    Namely, 
    $\mathbf{J}_{\textrm{C}}$ breaks time-reversal symmetry, but $\mathbf{J}_{\textrm{S}}$ does not.
  }
\end{figure*}

This difference between $\sigma_{yx}^{\textrm{S}}$ and $\sigma_{yx}^{\textrm{C}}$
comes from the difference in TRS. 
Under the time-reversal operation $T_{\textrm{rev}}$,
time $t$, momentum $\mathbf{k}$, and spin $\sigma$ are changed as follows: 
$(t,\mathbf{k},\sigma)\rightarrow (-t,-\mathbf{k},-\sigma)$,
where $-\sigma= \downarrow$ or $\uparrow$ for $\sigma= \uparrow$ or $\downarrow$,
respectively.
The spin current and charge current are expressed
as 
$\mathbf{J}_{\textrm{S}}=\frac{1}{2}(\mathbf{J}_{\uparrow}-\mathbf{J}_{\downarrow})$
and $\mathbf{J}_{\textrm{C}}=(-e)(\mathbf{J}_{\uparrow}+\mathbf{J}_{\downarrow})$,
where $\mathbf{J}_{\uparrow}$ and $\mathbf{J}_{\downarrow}$ are
the contributions from the spin-up and spin-down electrons,
respectively.
Thus,
$(\mathbf{J}_{\textrm{S}},\mathbf{J}_{\textrm{C}})\rightarrow (\mathbf{J}_{\textrm{S}}, -\mathbf{J}_{\textrm{C}})$
is obtained as a result of $T_{\textrm{rev}}$ 
because $(\mathbf{J}_{\uparrow},\mathbf{J}_{\downarrow})\rightarrow (-\mathbf{J}_{\downarrow},-\mathbf{J}_{\uparrow})$
is satisfied under $T_{\textrm{rev}}$ (Figs. \ref{fig3}a and \ref{fig3}b).
(This is the reason why TRS is broken in the AHE and not broken in the SHE.)
Meanwhile,
the right- and left-circularly polarized light fields
are connected by $T_{\textrm{rev}}$
because $\mathbf{A}_{\textrm{RCP}}(-t)=\mathbf{A}_{\textrm{LCP}}(t)$.
Namely,
replacing $\mathbf{A}_{\textrm{LCP}}(t)$ by $\mathbf{A}_{\textrm{RCP}}(t)$ 
corresponds to applying $T_{\textrm{rev}}$.
Thus,
the helicity-independent $\sigma_{yx}^{\textrm{S}}$
and the helicity-dependent $\sigma_{yx}^{\textrm{C}}$
result from $\mathbf{J}_{\textrm{S}}\rightarrow \mathbf{J}_{\textrm{S}}$
and $\mathbf{J}_{\textrm{C}}\rightarrow -\mathbf{J}_{\textrm{C}}$, respectively, 
under $T_{\textrm{rev}}$.

  The same helicity dependences
  hold in many periodically driven multiorbital metals.
  The spin current and charge current are of the same form
  for some transition metals (e.g., Pt and Au)~\cite{Kon-SHE-Pt}
  and transition-metal oxides.
  Then, the similar SHE and AHE can be realized
  using circularly polarized light.
  Thus,
  the above arguments are applicable to many transition-metal oxides and transition metals
  driven by circularly polarized light. 

\hspace{-13pt}
\textbf{SOC-dependent $\sigma_{yx}^{\textrm{S}}$ and SOC-independent $\sigma_{yx}^{\textrm{C}}$}

\begin{figure}
  \includegraphics[width=86mm]{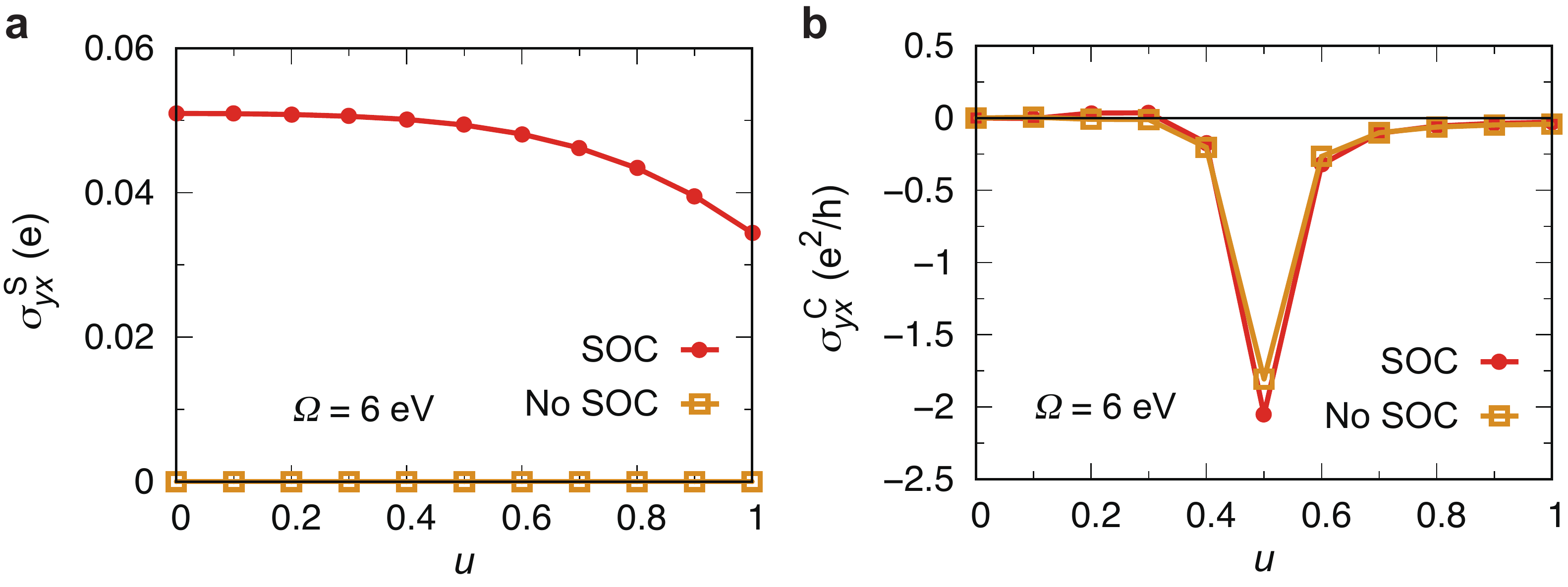}
  \caption{\label{fig4}
    \textbf{Spin-orbit coupling dependences of
      the spin Hall and anomalous Hall conductivities.}
    \textbf{a}, \textbf{b}
    The dependences of the spin Hall and anomalous Hall conductivities
    $\sigma_{yx}^{\textrm{S}}$ and $\sigma_{yx}^{\textrm{C}}$
    on the dimensionless quantity $u=eA_{0}$
    in the case of left-circularly polarized light
    at $\Omega=6$ eV with and without spin-orbit coupling.
    Here $\Omega$ is the frequency of light. 
    The red and yellow curves correspond to those
    with and without spin-orbit coupling, respectively. 
  }
\end{figure}

There is another difference between $\sigma_{yx}^{\textrm{S}}$ and $\sigma_{yx}^{\textrm{C}}$.
Figure \ref{fig4}a compares the $u$ dependence of $\sigma_{yx}^{\textrm{S}}$ with SOC
to that without SOC.
In the absence of SOC, $\sigma_{yx}^{\textrm{S}}=0$.
This is because there is no spin-dependent term in the Hamiltonian except for SOC.
The spin-dependent term, such as SOC, is needed to
obtain the finite difference between the spin-up and spin-down electron currents.
Meanwhile,
the $u$ dependence of $\sigma_{yx}^{\textrm{C}}$ with SOC is almost the same as
that without SOC (Fig. \ref{fig4}b).
This is because
a spin-independent electron current can be generated
by using the kinetic energy terms
with the Peierls phase factors~\cite{Oka-PRB,Mikami}
and a multiorbital mechanism~\cite{Kon-AHE} using SOC
  does not contribute to $\sigma_{yx}^{\textrm{C}}$
  in the presence of spin degeneracy,
  which is not lifted by the Peierls phase factors.
  Note that in periodically driven systems,
  $\sigma_{yx}^{\textrm{C}}$ can be finite
  even without orbital degrees of freedom~\cite{Oka-PRB,Mikami}
  because the Peierls phase factors can lead to
  the terms odd with respect to momentum in
  the energy dispersion (see Supplementary Note 2).

These results suggest that
in periodically driven multiorbital metals, 
SOC is vital for the SHE, whereas it is unnecessary for the AHE.
This suggestion may be valid
as long as the effects of the driving field can be treated as the Peierls phase factors
and there is no magnetic order.
In addition,
this is distinct from the property of non-driven multiorbital metals
where 
SOC is vital for
the SHE and AHE~\cite{Karplus-Luttinger,Kon-SHE-Pt,Kon-SHE-Ru,AHE-review,SHE-review,Kon-AHE}.
In contrast, the multiorbital nature is required for the SHE
of periodically driven systems,
whereas it is unnecessary for the AHE.
  
\hspace{-13pt}
\textbf{Implications and experimental realization}

We discuss some implications of our results.
First,
the difference between the helicity dependences
of $\sigma_{yx}^{\textrm{S}}$ and $\sigma_{yx}^{\textrm{C}}$
can be used to distinguish the spin current and charge current without ambiguity.
Since that difference results from the symmetry of $T_{\textrm{rev}}$,
the same helicity dependences should hold in many periodically driven systems.
In addition,
the similar arguments enable us to distinguish two currents,
one of which breaks TRS (and the other does not),
in not only Hall effects, but also other transport phenomena.
Thus,
our results have revealed the core physics discipline
about the relations between TRS and transport properties of periodically driven systems. 
Then, our theory can be extended to the SHE and AHE of other multiorbital metals
and other transport phenomena.
For example,
a combination of it and first-principles calculations
enables us to systematically search the SHE and AHE of periodically driven multiorbital metals.
Thus, 
our results provide the first step towards
the Floquet engineering of spintronics phenomena, including the SHE,
of periodically driven multiorbital metals.

Finally, we comment on experimental realization.
In our theory,
interaction effects and heating effects are neglected. 
For Sr$_{2}$RuO$_{4}$,
electron-electron interactions cause 
the orbital-dependent damping and
mass enhancement~\cite{Sr2RuO4-review,NA-FLEX}.
Since these effects are quantitative~\cite{Kon-SHE-Ru},
the interaction effects may not change our results at least qualitatively. 
The differences in the helicity dependence and the SOC dependence will hold 
because those interaction effects do not break TRS.
In general,
the periodic driving makes the system to heat up~\cite{Heating-iso}. 
However,
for the periodically driven open system, such as our system,
a nonequilibrium steady state can be realized
due to $\Gamma$~\cite{Heating-open,Tsuji,Mikami}
at times larger than
$\tau(=\hbar/2\Gamma)\approx 11\textrm{fs} =O(10\textrm{fs})$.
In fact, the AHE predicted theoretically
in a periodically driven open system~\cite{Oka-PRB} is
experimentally realized~\cite{Light-AHE-exp1,Light-AHE-exp2}.
For Sr$_{2}$RuO$_{4}$, in which $a_{\textrm{lc}}\approx 0.39$ nm~\cite{Sr2RuO4-review}, 
$u(=ea_{\textrm{lc}}A_{0})=0.3$ at $\Omega=2$, $4$, or $6$ eV
corresponds to $E_{0}=A_{0}/\Omega\approx 15$, $31$, or $46$ MVcm$^{-1}$, respectively. 
Since the pump field of the order of $10$ MVcm$^{-1}$ is experimentally accessible~\cite{Iwai-review}, 
we conclude that
the predicted properties of 
$\sigma_{yx}^{\textrm{S}}$ and $\sigma_{yx}^{\textrm{C}}$
could be observed in the pump-probe measurements
of the SHE and AHE in periodically driven Sr$_{2}$RuO$_{4}$. 

\section{Methods}

\subsection{Tight-binding Hamiltonian with SOC}

We have chosen the following tight-binding Hamiltonian for $t_{2g}$-orbital electrons
as $H_{\textrm{s}}(t)$:
\begin{align}
  H_{\textrm{s}}(t)
  &=\sum_{i,j}\sum_{a,b=d_{yz},d_{zx},d_{xy}}\sum_{\sigma=\uparrow,\downarrow}
  [t_{ij}^{ab}(t)-\mu\delta_{i,j}\delta_{a,b}]c_{ia\sigma}^{\dagger}c_{jb\sigma}\notag\\
  &+\sum_{i}\sum_{a,b=d_{yz},d_{zx},d_{xy}}\sum_{\sigma,\sigma^{\prime}=\uparrow,\downarrow}
  \xi_{ab}^{\sigma\sigma^{\prime}}c_{ia\sigma}^{\dagger}c_{ib\sigma^{\prime}},\label{eq:Hs-start}
\end{align}
where
$t_{ij}^{ab}(t)$'s are the hopping integrals with the Peierls phase factors due to $\mathbf{A}(t)$, 
$t_{ij}^{ab}(t)=t_{ij}^{ab}e^{-ie(\mathbf{R}_{i}-\mathbf{R}_{j})\cdot\mathbf{A}(t)}$,
and $\xi_{ab}^{\sigma\sigma^{\prime}}$
is the coupling constant of the SOC for $t_{2g}$-orbital electrons.
The finite elements of $\xi_{ab}^{\sigma\sigma^{\prime}}=(\xi_{ba}^{\sigma^{\prime}\sigma})^{\ast}$
are given by
$\xi_{d_{yz} d_{zx}}^{\uparrow\uparrow}=\xi_{d_{zx} d_{xy}}^{\uparrow\downarrow}=i\xi/2$,
$\xi_{d_{yz} d_{xy}}^{\uparrow\downarrow}=-\xi/2$,
$\xi_{d_{xy}d_{yz}}^{\uparrow\downarrow}=\xi/2$, and 
$\xi_{d_{xy} d_{zx}}^{\uparrow\downarrow}=\xi_{d_{yz} d_{zx}}^{\downarrow\downarrow}=-i\xi/2$.
By using the Fourier coefficients of the operators,
we can write Eq. (\ref{eq:Hs-start}) as Eq. (\ref{eq:Hs})
with Eq. (\ref{eq:ebar}), in which 
$\epsilon_{ab}(\mathbf{k},t)$ is given by
$\epsilon_{ab}(\mathbf{k},t)=\sum_{j}t_{ij}^{ab}(t)e^{-i\mathbf{k}\cdot(\mathbf{R}_{i}-\mathbf{R}_{j})}$.

\subsection{B\"{u}ttiker-type heat bath}

$H_{\textrm{sb}}$ and $H_{\textrm{b}}$ in Eq. (\ref{eq:Htot}) are given by
\begin{align}
  &H_{\textrm{sb}}
  =\sum_{i}\sum_{p}
  \sum_{a=d_{yz},d_{zx},d_{xy}}\sum_{\sigma=\uparrow,\downarrow}
  V_{pa\sigma}(c_{ia\sigma}^{\dagger}b_{ip}+b_{ip}^{\dagger}c_{ia\sigma}),\label{eq:Hsb}\\
  &H_{\textrm{b}}
  =\sum_{i}\sum_{p}
  (\epsilon_{p}-\mu_{\textrm{b}})b_{ip}^{\dagger}b_{ip},\label{eq:Hb}
\end{align}
where
$b_{ip}$ and $b_{ip}^{\dagger}$ are the annihilation and creation operators,
respectively, of a bath's fermion at site $i$ for mode $p$,
$V_{pa\sigma}$ is the coupling constant,
and 
$\epsilon_{p}$ and $\mu_{\textrm{b}}$
are the energy and chemical potential of a bath's fermion.
Note that $\mu_{\textrm{b}}$ is chosen in order that there is no current between
the system and bath. 
The heat bath is supposed to be in equilibrium at temperature $T_{\textrm{b}}$. 
The main effect of the heat bath is 
the damping appearing in electron Green's functions~\cite{Tsuji,Mikami}. 

\subsection{Charge current and spin current operators}

We derive the charge current and spin current operators
using the continuity equations.
Theories using these operators derived in that way
succeed in describing the SHE
observed in non-driven multiorbital metals~\cite{Kon-SHE-Pt,SHE-d-exp}. 

First,
we derive the charge current operator $\mathbf{J}_{\textrm{C}}(t)$. 
$\mathbf{J}_{\textrm{C}}(t)$ is supposed to
satisfy the continuity equation~\cite{Mahan},
\begin{align}
  \frac{d\rho_{j}(t)}{dt}+\nabla\cdot \mathbf{j}_{j}^{(\textrm{C})}(t)=0,\label{eq:cont-C}
\end{align}
where $\rho_{j}(t)=(-e)\sum_{a}\sum_{\sigma}c_{ja\sigma}^{\dagger}(t)c_{ja\sigma}(t)$
and $\sum_{j}\mathbf{j}_{j}^{(\textrm{C})}(t)=\mathbf{J}_{\textrm{C}}(t)$.
Using Eq. (\ref{eq:cont-C}),
we have
\begin{align}
  \sum_{j}\mathbf{R}_{j}\frac{d\rho_{j}(t)}{dt}
  =-\sum_{j}\mathbf{R}_{j}\nabla\cdot \mathbf{j}_{j}^{(\textrm{C})}(t)
  =\mathbf{J}_{\textrm{C}}(t),\label{eq:cont-C-next}
\end{align}
where we have omitted the surface contributions. 
By combining it with the Heisenberg equation,
we can write Eq. (\ref{eq:cont-C-next}) as
\begin{align}
  \mathbf{J}_{\textrm{C}}(t)=i[H_{\textrm{s}}(t),\sum_{j}\mathbf{R}_{j}\rho_{j}(t)].
\end{align}
(Note that
there is no contribution from $H_{\textrm{sb}}$
because the bath's chemical potential is chosen
in order that there is no current between the system and bath.)
After some calculations,
we obtain
\begin{align}
  \mathbf{J}_{\textrm{C}}(t)
  =&i\sum_{i,j}\sum_{a,b}\sum_{\sigma}(-e)t_{ij}^{ab}(t)(\mathbf{R}_{j}-\mathbf{R}_{i})
  c_{ia\sigma}^{\dagger}(t)c_{jb\sigma}(t)\notag\\
  =&-e\sum_{\mathbf{k}}\sum_{a,b}\sum_{\sigma}
  \frac{\partial \epsilon_{ab}(\mathbf{k},t)}{\partial \mathbf{k}}
  c_{\mathbf{k} a\sigma}^{\dagger}(t)c_{\mathbf{k} b\sigma}(t).
\end{align}

Similarly, we derive
the spin current operator $\mathbf{J}_{\textrm{S}}(t)$.
We suppose that
$\mathbf{J}_{\textrm{S}}(t)$ satisfies
\begin{align}
  \frac{dS_{j}^{z}(t)}{dt}+\nabla\cdot \mathbf{j}_{j}^{(\textrm{S})}(t)=0,\label{eq:cont-S}
\end{align}
where $S_{j}^{z}(t)=\sum_{a}\sum_{\sigma}\frac{1}{2}\textrm{sgn}(\sigma)c_{ja\sigma}^{\dagger}(t)c_{ja\sigma}(t)$
and $\sum_{j}\mathbf{j}_{j}^{(\textrm{S})}(t)=\mathbf{J}_{\textrm{S}}(t)$.
In a way similar to the derivation of $\mathbf{J}_{\textrm{C}}(t)$,
$\mathbf{J}_{\textrm{S}}(t)$ is given by
\begin{align}
  \mathbf{J}_{\textrm{S}}(t)
  =&i[H_{\textrm{s}}(t),\sum_{j}\mathbf{R}_{j}S_{j}^{z}(t)]\notag\\
  =&\frac{1}{2}\sum_{\mathbf{k}}\sum_{a,b}\sum_{\sigma}
  \textrm{sgn}(\sigma)\frac{\partial \epsilon_{ab}(\mathbf{k},t)}{\partial \mathbf{k}}
  c_{\mathbf{k} a\sigma}^{\dagger}(t)c_{\mathbf{k} b\sigma}(t).
\end{align}

\subsection{Anomalous-Hall and spin-Hall conductivities as functions of time}

We express $\sigma_{yx}^{\textrm{C}}(t,t^{\prime})$ and $\sigma_{yx}^{\textrm{S}}(t,t^{\prime})$
in terms of the electron Green's functions.
Using Eq. (\ref{eq:JC-JS}),
we have
\begin{align}
  \langle j_{\textrm{C}}^{y}(t)\rangle
  &=\frac{-i}{N}\sum_{\mathbf{k}}\sum_{a,b}\sum_{\sigma}
  v_{ab\sigma}^{(\textrm{C})y}(\mathbf{k},t)G_{b\sigma a\sigma}^{<}(\mathbf{k};t,t),\label{eq:jCy-start}\\
  \langle j_{\textrm{S}}^{y}(t)\rangle
  &=\frac{-i}{N}\sum_{\mathbf{k}}\sum_{a,b}\sum_{\sigma}
  v_{ab\sigma}^{(\textrm{S})y}(\mathbf{k},t)G_{b\sigma a\sigma}^{<}(\mathbf{k};t,t),\label{eq:jSy-start}
\end{align}
where $G_{b\sigma^{\prime} a\sigma}^{<}(\mathbf{k};t,t^{\prime})$ is
the lesser Green's function~\cite{Mahan,Keldysh,Keldysh-review,Kadanoff-Baym},
\begin{align}
  G_{b\sigma^{\prime} a\sigma}^{<}(\mathbf{k};t,t^{\prime})
  =i\langle c_{\mathbf{k} a\sigma}^{\dagger}(t^{\prime})c_{\mathbf{k} b\sigma^{\prime}}(t)\rangle.
\end{align}
By substituting Eqs. (\ref{eq:jCy-start}) and (\ref{eq:jSy-start})
into Eq. (\ref{eq:AHC-SHC-start}),
we can express $\sigma_{yx}^{\textrm{C}}(t,t^{\prime})$ and $\sigma_{yx}^{\textrm{S}}(t,t^{\prime})$
as follows: 
\begin{align}
  \sigma_{yx}^{\textrm{C}}(t,t^{\prime})
  &=\sigma_{yx}^{\textrm{C}(1)}(t,t^{\prime})+\sigma_{yx}^{\textrm{C}(2)}(t,t^{\prime}),\\
  \sigma_{yx}^{\textrm{S}}(t,t^{\prime})
  &=\sigma_{yx}^{\textrm{S}(1)}(t,t^{\prime})+\sigma_{yx}^{\textrm{S}(2)}(t,t^{\prime}),
\end{align}
where 
\begin{align}
  &\sigma_{yx}^{\textrm{Q}(1)}(t,t^{\prime})
  =\frac{-1}{\omega N}\sum_{\mathbf{k}}\sum_{a,b}\sum_{\sigma}
  \frac{\delta v_{ab\sigma}^{(\textrm{Q})y}(\mathbf{k},t)}{\delta A_{\textrm{prob}}^{x}(t^{\prime})}
  G_{b\sigma a\sigma}^{<}(\mathbf{k};t,t),\\
  &\sigma_{yx}^{\textrm{Q}(2)}(t,t^{\prime})
  =\frac{-1}{\omega N}\sum_{\mathbf{k}}\sum_{a,b}\sum_{\sigma}
  v_{ab\sigma}^{(\textrm{Q})y}(\mathbf{k},t)
  \frac{\delta G_{b\sigma a\sigma}^{<}(\mathbf{k};t,t)}{\delta A_{\textrm{prob}}^{x}(t^{\prime})}.
  \label{eq:sigPM-start}
\end{align}
Then,
using the Dyson equation of Green's functions
and the Langreth rule~\cite{Mikami,Keldysh-review},
we obtain
\begin{align}
  &\frac{\delta G_{b\sigma a\sigma}^{<}(\mathbf{k};t,t)}{\delta A_{\textrm{prob}}^{x}(t^{\prime})}\notag\\
  =&-\sum_{c,d}\sum_{\sigma^{\prime}}v_{cd\sigma^{\prime}}^{(\textrm{C})x}(\mathbf{k},t^{\prime})
  \Bigl[
    G_{b\sigma c\sigma^{\prime}}^{\textrm{R}}(\mathbf{k};t,t^{\prime})
    G_{d\sigma^{\prime}a\sigma}^{<}(\mathbf{k};t^{\prime},t)\notag\\
    &\ \ \ \ \ \ \ \ \ \ \ \ \ \ \ \ \ \ \ 
    +G_{b\sigma c\sigma^{\prime}}^{<}(\mathbf{k};t,t^{\prime})
    G_{d\sigma^{\prime}a\sigma}^{\textrm{A}}(\mathbf{k};t^{\prime},t)
    \Bigr],\label{eq:rewrite}
\end{align}
where $G_{a\sigma b\sigma^{\prime}}^{\textrm{R}}(\mathbf{k};t,t^{\prime})$
and $G_{a\sigma b\sigma^{\prime}}^{\textrm{A}}(\mathbf{k};t,t^{\prime})$
are the retarded and advanced Green's
functions~\cite{Mahan,Keldysh,Keldysh-review,Kadanoff-Baym}, respectively,
\begin{align}
  G_{a\sigma b\sigma^{\prime}}^{\textrm{R}}(\mathbf{k};t,t^{\prime})
  &=-i\theta(t-t^{\prime})\langle \{c_{\mathbf{k} a\sigma}(t),c_{\mathbf{k} b\sigma^{\prime}}^{\dagger}(t^{\prime})\}\rangle,\\
  G_{a\sigma b\sigma^{\prime}}^{\textrm{A}}(\mathbf{k};t,t^{\prime})
  &=i\theta(t^{\prime}-t)\langle \{c_{\mathbf{k} a\sigma}(t),c_{\mathbf{k} b\sigma^{\prime}}^{\dagger}(t^{\prime})\}\rangle.
\end{align}
Combining Eq. (\ref{eq:rewrite}) with Eq. (\ref{eq:sigPM-start}),
we have
\begin{align}
  \sigma_{yx}^{\textrm{Q}(2)}(t,t^{\prime})
  &=\frac{1}{\omega N}\sum_{\mathbf{k}}\sum_{a,b,c,d}\sum_{\sigma,\sigma^{\prime}}
  v_{ab\sigma}^{(\textrm{Q})y}(\mathbf{k},t)v_{cd\sigma^{\prime}}^{(\textrm{C})x}(\mathbf{k},t^{\prime})\notag\\
  &\times
  \Bigl[G_{b\sigma c\sigma^{\prime}}^{\textrm{R}}(\mathbf{k};t,t^{\prime})
    G_{d\sigma^{\prime} a\sigma}^{<}(\mathbf{k};t^{\prime},t)\notag\\
    &+G_{b\sigma c\sigma^{\prime}}^{<}(\mathbf{k};t,t^{\prime})
    G_{d\sigma^{\prime} a\sigma}^{\textrm{A}}(\mathbf{k};t^{\prime},t)
    \Bigr].
\end{align}

\subsection{Dyson equation of Green's functions}

The Green's functions of our periodically driven system
are determined from the Dyson equation in a matrix form: 
\begin{align}
  G=G_{0}+G_{0}\Sigma G,\label{eq:Dyson}
\end{align}
where $G$, $G_{0}$, and $\Sigma$ are the matrices of
the Green's functions with $H_{\textrm{sb}}$,
those without $H_{\textrm{sb}}$,
and the self-energies due to the second-order perturbation of $H_{\textrm{sb}}$,
respectively,
\begin{align}
  G=\left(
  \begin{array}{@{\,}cc@{\,}}
    G^{\textrm{R}} & G^{\textrm{K}}\\[3pt]
    0 & G^{\textrm{A}}
  \end{array}
  \right),
  G_{0}=\left(
  \begin{array}{@{\,}cc@{\,}}
    G^{\textrm{R}}_{0} & G^{\textrm{K}}_{0}\\[3pt]
    0 & G^{\textrm{A}}_{0}
  \end{array}
  \right),
  \Sigma=\left(
  \begin{array}{@{\,}cc@{\,}}
    \Sigma^{\textrm{R}} & \Sigma^{\textrm{K}}\\[3pt]
    0 & \Sigma^{\textrm{A}}
  \end{array} 
  \right)\label{eq:Keldysh-rep}.
\end{align}
The superscripts R, A, and K denote
the retarded, advanced, and Keldysh components, respectively. 
For example,
the matrix $G^{\textrm{R}}$
as a function of $\mathbf{k}$ and $\omega$ 
is given by 
$G^{\textrm{R}}=([G^{\textrm{R}}_{a\sigma b\sigma^{\prime}}(\mathbf{k},\omega)]_{mn})$
for $a,b=d_{yz},d_{zx},d_{xy}$, $\sigma,\sigma^{\prime}=\uparrow,\downarrow$,
and $m,n=-\infty,\cdots,0,\cdots,\infty$.
The retarded, advanced, and Keldysh components are related to the lesser one
through the identity,
such as
\begin{align}
  G^{<}=\frac{1}{2}(G^{\textrm{K}}-G^{\textrm{R}}+G^{\textrm{A}}).\label{eq:G^<-relation}
\end{align}
By treating the effects of $H_{\textrm{sb}}$ in the second-order perturbation theory,
we can express the retarded, advanced, and Keldysh self-energies as follows:
\begin{align}
  [\Sigma^{\textrm{R}}_{a\sigma b\sigma^{\prime}}(\mathbf{k},\omega)]_{mn}
  &=
  -i\delta_{m,n}\delta_{a,b}\delta_{\sigma,\sigma^{\prime}}\Gamma,\label{eq:Sig^R}\\
  [\Sigma^{\textrm{A}}_{a\sigma b\sigma^{\prime}}(\mathbf{k},\omega)]_{mn}
  &=+i\delta_{m,n}\delta_{a,b}\delta_{\sigma,\sigma^{\prime}}\Gamma,\label{eq:Sig^A}\\
  [\Sigma^{\textrm{K}}_{a\sigma b\sigma^{\prime}}(\mathbf{k},\omega)]_{mn}
  &=-2i\delta_{m,n}\delta_{a,b}\delta_{\sigma,\sigma^{\prime}}\Gamma
  \tanh\frac{\omega+m\Omega}{2T_{\textrm{b}}}.\label{eq:Sig^K}
\end{align}
In deriving them, 
we have omitted the real parts 
and replaced
$\pi\sum_{p}V_{pa\sigma}V_{pb\sigma^{\prime}}\delta(\omega+m\Omega-\epsilon_{p}+\mu_{\textrm{b}})$
by $\Gamma\delta_{a,b}\delta_{\sigma,\sigma^{\prime}}$ for simplicity.
Such simplification may be sufficient
because the main effect of $H_{\textrm{sb}}$ is
the relaxation towards the nonequilibrium steady state
due to the damping~\cite{Tsuji,Mikami}.
Then,
using the matrix relation $G^{-1}G=1$ and Eq. (\ref{eq:Keldysh-rep}),
we have
\begin{align}
  &(G^{\textrm{R}})^{-1}=(G^{-1})^{\textrm{R}},\label{eq:G^R-inv}\\
  &(G^{\textrm{A}})^{-1}=(G^{-1})^{\textrm{A}},\label{eq:G^A-inv}\\
  &G^{\textrm{K}}=-G^{\textrm{R}}(G^{-1})^{\textrm{K}}G^{\textrm{A}},\label{eq:G^K-inv}
\end{align}
where
\begin{align}
  G^{-1}=\left(
  \begin{array}{@{\,}cc@{\,}}
    (G^{-1})^{\textrm{R}} & (G^{-1})^{\textrm{K}}\\[3pt]
    0 & (G^{-1})^{\textrm{A}}
  \end{array}
  \right).
\end{align}
Therefore, 
the retarded and advanced Green's functions with $H_{\textrm{sb}}$
are obtained by calculating the inverse matrices of
$(G^{-1})^{\textrm{R}}$ and $(G^{-1})^{\textrm{A}}$, respectively,
\begin{align}
  [(G^{-1})^{\textrm{R}}_{a\sigma b\sigma^{\prime}}(\mathbf{k},\omega)&]_{mn}
  =(\omega+\mu+m\Omega+i\Gamma)\delta_{m,n}\delta_{a,b}\delta_{\sigma,\sigma^{\prime}}\notag\\
  &-\xi_{ab}^{\sigma\sigma^{\prime}}\delta_{m,n}
  -[\epsilon_{ab}(\mathbf{k})]_{mn}\delta_{\sigma,\sigma^{\prime}},\\
  [(G^{-1})^{\textrm{A}}_{a\sigma b\sigma^{\prime}}(\mathbf{k},\omega)&]_{mn}
  =(\omega+\mu+m\Omega-i\Gamma)\delta_{m,n}\delta_{a,b}\delta_{\sigma,\sigma^{\prime}}\notag\\
  &-\xi_{ab}^{\sigma\sigma^{\prime}}\delta_{m,n}
  -[\epsilon_{ab}(\mathbf{k})]_{mn}\delta_{\sigma,\sigma^{\prime}},
\end{align}
where
\begin{align}
  [\epsilon_{ab}(\mathbf{k})]_{mn}
  =\int_{0}^{T}\frac{dt}{T}e^{i(m-n)\Omega t}\epsilon_{ab}(\mathbf{k},t).
\end{align}
The expressions of $[\epsilon_{ab}(\mathbf{k})]_{mn}$ for our model
  are provided in Supplementary Note 2;
  as shown there, $[\epsilon_{ab}(\mathbf{k})]_{mn}$
  includes the Bessel functions of the first kind as a function of $u=eA_{0}$.
After obtaining these Green's functions,
we can obtain 
the Keldysh Green's function with $H_{\textrm{sb}}$ 
by combining Eq. (\ref{eq:G^K-inv}) with the following equation:
\begin{align}
  [(G^{-1})^{\textrm{K}}_{a\sigma b\sigma^{\prime}}(\mathbf{k},\omega)]_{mn}
  =2i\Gamma\delta_{m,n}\delta_{a,b}\delta_{\sigma,\sigma^{\prime}}\Gamma
  \tanh\frac{\omega+m\Omega}{2T_{\textrm{b}}}.
\end{align}
We finally obtain the lesser Green's function with $H_{\textrm{sb}}$
using the three Green's functions obtained and Eq. (\ref{eq:G^<-relation}).

\subsection{Numerical calculations}

We numerically calculated Eq. (\ref{eq:sig^Q}) for Q $=$ C or S,
$\sigma_{yx}^{\textrm{C}}$ or $\sigma_{yx}^{\textrm{S}}$,
in the following way.
The momentum summation was calculated
by dividing the Brillouin zone into a $N_{x}\times N_{y}$ mesh
and setting $N_{x}=N_{y}=100$.
The frequency integral was done 
by using
$\int_{-\Omega/2}^{\Omega/2}d\omega^{\prime}F(\omega^{\prime})
\approx \sum_{s=0}^{W-1}\Delta\omega^{\prime} F(\omega^{\prime}_{s})$,
where 
$\omega^{\prime}_{s}=-\Omega/2+s\Delta\omega^{\prime}$
and $\omega^{\prime}_{W}=\Omega/2$,
and setting $\Delta\omega^{\prime}=0.005$ eV. 
The frequency derivatives of the Green's functions was approximated
by using
$\frac{\partial F(\omega^{\prime})}{\partial \omega^{\prime}}
\approx \frac{F(\omega^{\prime}+\Delta\omega^{\prime})-F(\omega^{\prime}-\Delta\omega^{\prime})}{2\Delta\omega^{\prime}}$. 
The summations over the Floquet indices, $\sum_{m,l,n,q=-\infty}^{\infty}$,
was replaced by $\sum_{m,l,n,q=-n_{\textrm{max}}}^{n_{\textrm{max}}}$,
and $n_{\textrm{max}}$ was fixed at $n_{\textrm{max}}=2$ for $\Omega=6$ and $4$ eV
or $n_{\textrm{max}}=3$ for $\Omega=2$ eV.

\section{Data availability}

The data that support the findings of this study
are available from
the corresponding author upon reasonable request.

\section{Code availability}

The code used in the numerical calculations is available from
the corresponding author upon reasonable request.

\section{Acknowledgments}

This work was supported by
JST CREST Grant No. JPMJCR1901, 
JSPS KAKENHI Grants No. JP22K03532, JP19K14664, and JP16K05459,
and MEXT Q-LEAP Grant No. JP-MXS0118067426.

\section{Author contributions}

N.A. conceived the project, formulated the theory, performed the numerical calculations,
and wrote the manuscript.
K.Y. supervised the project.
All authors discussed the results and commented on the manuscript.

\section{Competing interests}

The authors declare no competing interests.

\section{Additional information}

\textbf{Supplementary Information} The online version
contains supplementary material. 

\textbf{Correspondence and requests for materials} should be addressed to N.A. 

\end{document}